\documentclass[iop,apjl,tighten]{emulateapj}
\usepackage{amssymb}
\usepackage{times}
\usepackage{epsfig}
\usepackage{graphics}
\usepackage{natbib}

\journalinfo{The Astrophysical Journal Letters, 717: L118-L121, 2010 July 10\hspace{3cm} doi:10.1088/2041-8205/717/2/L118}

\begin{document}

\title{GeV breaks in blazars as a result of gamma-ray absorption within the broad-line region} 

\shorttitle{GeV breaks in blazars as a result of gamma-ray absorption within BLR}
\shortauthors{Poutanen \& Stern}

\author{Juri Poutanen\altaffilmark{1} and Boris Stern\altaffilmark{1,2,3}} 

\affil{$^1$Astronomy Division, Department of Physics, P.O.Box 3000, 
90014 University of Oulu, Finland; juri.poutanen@oulu.fi \\
$^2$Institute for Nuclear Research, Russian Academy of Sciences, Prospekt 60-letiya Oktyabrya 7a, Moscow 117312, Russia;
boris.stern@gmail.com\\
$^3$Astro Space Center, Lebedev Physical Institute, Profsoyuznaya 84/32,  Moscow 117997, Russia}

\submitted{Received 2010 May 20; accepted 2010 June 8; published 2010 June 21}

\begin{abstract} 
\noindent 
Spectra of the brightest blazars detected by the {\em Fermi Gamma-ray Space Telescope} Large Area Telescope
cannot be described by a simple power law model. 
A much better description is obtained with a broken power law, with the break energies of a few GeV.   
We show here that the sharpness and the position of the breaks can be well reproduced by absorption of $\gamma$-rays via 
photon--photon pair production on  He\,{\sc ii}  Lyman recombination continuum and lines. 
This implies that the blazar zone lies inside the region of the highest ionization of the broad-line region (BLR)
within a light-year from a super-massive black hole. 
The observations of $\gamma$-ray spectral breaks open a way of studying the BLR photon field in the extreme-UV/soft X-rays, 
which are otherwise hidden from our view. 
\end{abstract}



\section{Introduction}

The {\em Fermi Gamma-ray Space Telescope} has detected more than a hundred of blazars in the 100 MeV--100 GeV range \citep{Abdo09_AGN} 
and allowed their detailed spectral studies with unprecedented accuracy. 
A surprising result is that the spectra of high-luminosity sources, flat-spectra radio quasars (FSRQs) and low-energy synchrotron peaked BL Lac objects, are much better described by a broken power law than by a simple powerlaw or any smoothly curved models \citep{Abdo09_3C454.3,Abdo10_blazars}. The break energies, mostly lying in the 2--10 GeV range (as measured in the object frame),  seem to be too small to be produced by the  $\gamma$-ray absorption due to photon--photon pair production in the broad-line region (BLR), as the strongest BLR line,  
Ly$\alpha$, absorbs $\gamma$-rays starting from only 25.6 GeV.  The observed spectral breaks are also too sharp and the change in the spectral indices are too large to be associated with the cooling or the Klein-Nishina effects \citep{GT09}. 

Calculations of the $\gamma$-ray absorption within the BLR are usually  limited to the contribution from lines in the observed optical band 
(e.g. \citealt{Liu06,Reimer07}) with a few exceptions \citep{TM09}.
Theoretical models, however, predict strong helium UV lines as well as metal lines in the soft X-rays \citep[see, e.g.][]{Krolik99}, because a 
typical quasar spectrum has a power law tail extending to the X-ray band \citep{Laor97}. In addition to the lines, there are 
strong recombination continua of hydrogen and He\,{\sc ii}  at 13.6 and 54.4 eV, respectively. 
These sharp,  line-like features cause jumps in the $\gamma$-ray opacity at $\sim$19.2 and 4.8 GeV. 
We propose here that the observed spectral breaks in blazars are produced by absorption on photons of these recombination continua. 

\begin{figure*}
\centerline{\epsfig{file= 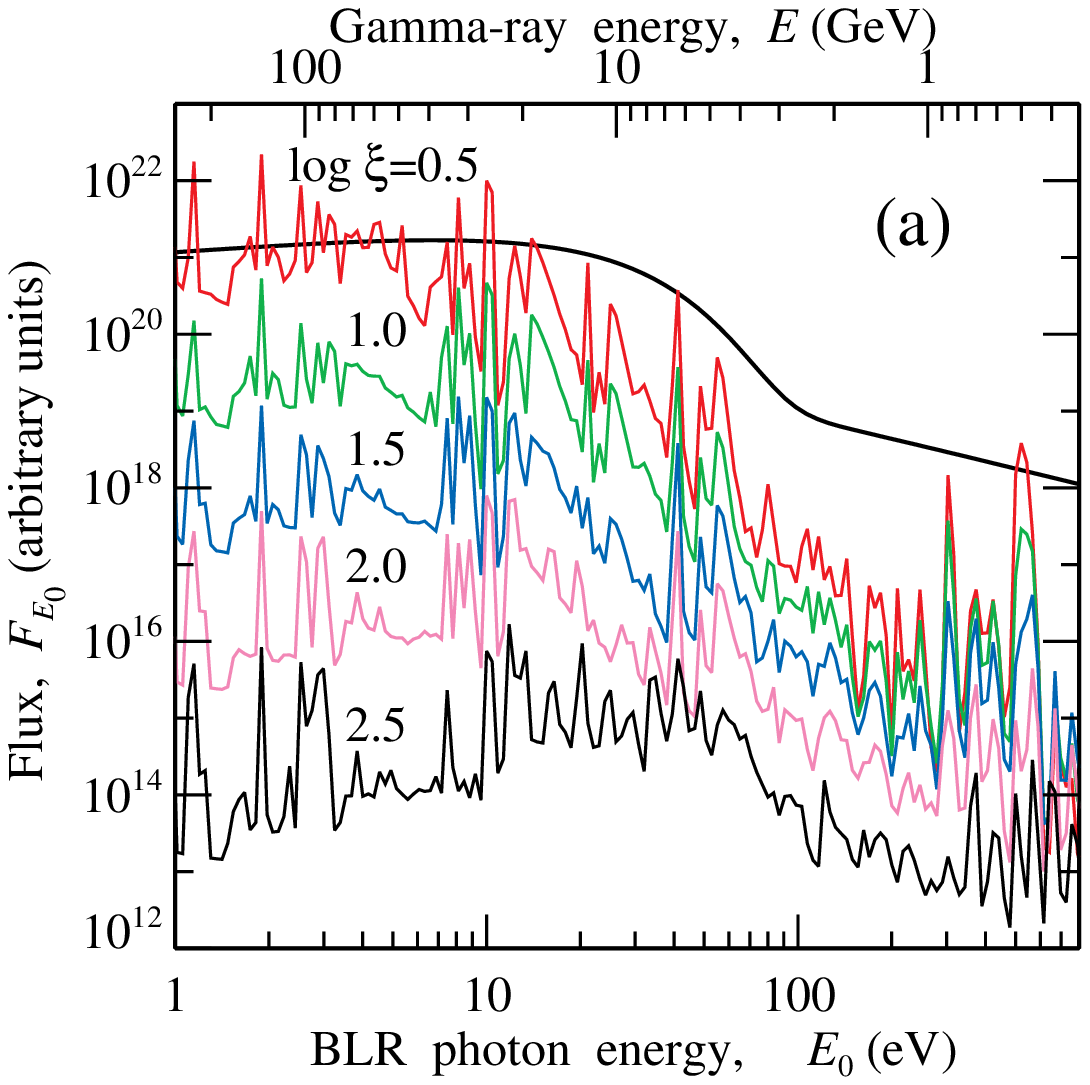,width=6.7cm}\hspace{2cm}\epsfig{file= 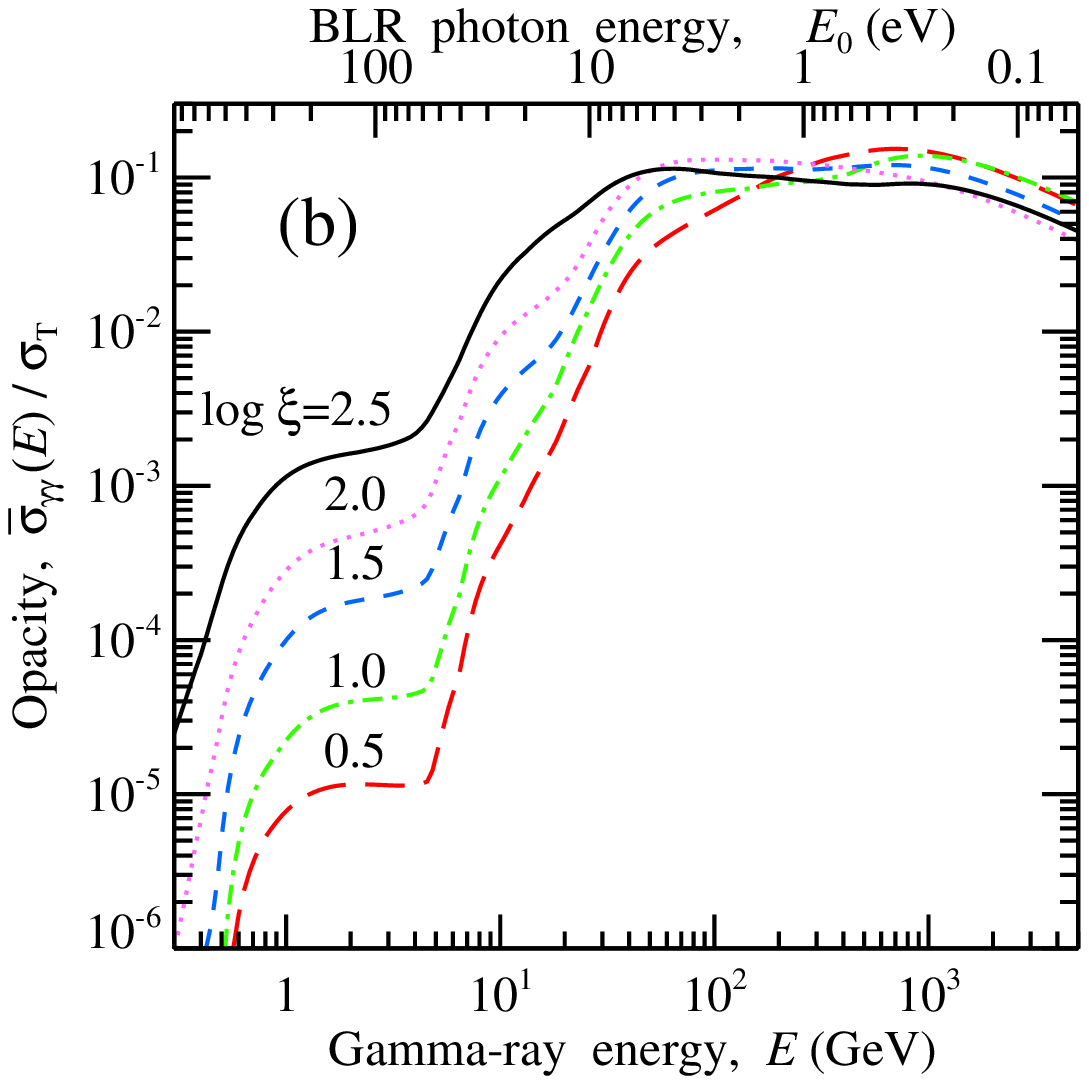,width=6.7cm}} 
\caption{ (a) Spectrum of the broad emission line region. The spectra for various  ionization parameters $\xi$
computed using a photoionization code {\sc xstar} version 2.2 \citep{KB01} 
are averaged on a logarithmic energy grid with the 5\%  
spectral resolution to show the most significant features dominating the total flux. 
The incident typical quasar spectrum \citep{Laor97}  shown by a smooth curve consists of a multicolor disk with the maximum temperature $10^5$ K and a power law tail of photon index $\Gamma=2$ extending to 100 keV with 10\% of the total luminosity.
(b) Average cross section (in units of Thomson cross section) for $\gamma$-rays due to photon--photon pair production on the BLR spectra shown in panel (a). Opacity produced by the disk radiation, direct or scattered outside the BLR clouds \citep[see e.g.][]{BL95}, is neglected here.
Jumps in opacity are clearly seen at energies corresponding to the strongest lines and recombination continua. 
The 0.3--0.5 GeV jump for large $\xi$ is produced by O\,{\sc viii} $\lambda\lambda$16--19  lines, the jump at low $\xi$ is due to the 
O\,{\sc vii}   $\lambda$22 line complex (see Table \ref{tbl:blr}). The jump at $\sim$5 GeV visible at all ionizations 
is due to the He\,{\sc ii} recombination continuum and Ly lines at 40--60 eV. 
The 20--30 GeV jump seen at low $\xi$ is produced by H\,{\sc i} Ly lines and continuum and the C\,{\sc iv} $\lambda$1549 line.  
}	
\label{fig:blr}
\end{figure*}

\section{Gamma-ray absorption within the broad-line region} 
\label{sec:blr}

\subsection{Emission From Broad-Line Region}

The  BLR around quasars is known to emit a number of  strong lines associated with very different ionization stages. 
Reverberation mapping of Seyfert galaxies and quasars using the C\,{\sc iv} $\lambda$1549 line 
showed that the size of the BLR scales with the luminosity as \citep{Kaspi07}
\begin{equation}\label{eq:rblr}
R_{\rm C\,IV,18} \approx 0.4  L_{47}^{1/2}.
\end{equation}
The estimated BLR size is, however, 2--3 times larger for the Balmer lines and 3 times smaller for 
high-ionization He\,{\sc ii}  $\lambda$1640 and N\,{\sc v} $\lambda$1240 lines \citep{Korista95,PW99}.
This argues in favor of strong radial ionization stratification of the BLR spanning over an order of magnitude in radius.  The BLR spectrum is thus very much dependent on the distance from the central ionizing source.

Using spectral synthesis code {\sc xstar} \citep[version 2.2,][]{KB01}, we   generated a
grid of photoionization models of BLR clouds assumed to be simple slabs of constant gas density and a clear view to the ionizing source. 
For simplicity, we  fix the  cloud column density  at $N_{\rm H}=10^{23}$ cm$^{-2}$ and 
vary the ionization parameter  $\xi=L/(r^2 n_{\rm H})$  from 10$^{0.5}$ to 10$^{2.5}$. 
We assume a gradual change of the cloud density as a function of distance from the central source, 
$n_{\rm H}\propto r^{-1}$. Taking $n_{\rm H}=10^{11.5}$ cm$^{-3}$ at the inner edge of the BLR, the distances $r$ to the ionizing source 
are between 0.03 ($\log \xi=2.5$) and 1 pc ($\log \xi=0.5$)  for a quasar luminosity $L=10^{47}$ erg s$^{-1}$. 
This BLR model is broadly consistent with the constraints from reverberation mapping \citep{KN99}. 
It is, of course, still an oversimplification, but this is enough to show the impact of BLR 
on propagation of the GeV photons. 

The quasar spectrum is taken as a sum of the standard multicolor accretion disk  plus a power law of total luminosity 10\% and extending to 100 keV \citep{Laor97}. The resulting BLR spectra are shown in Figure \ref{fig:blr}(a). 
Their absolute normalization  depend on the covering fraction of the BLR clouds. If  it does not vary dramatically over distance, 
the radiation field at a given radius is dominated by the locally produced photons corresponding to a specific ionization parameter. 
As our aim is not the study of the BLR emission, but its effect on the high-energy photon propagation, we smooth 
the spectrum to show the most significant features (see Table \ref{tbl:blr}). 
In the high-ionization zone at small distances to the central source, 
He\,{\sc ii} Ly lines  and recombination continuum  at  40--60 eV are relatively strong. 
There are also significant soft X-ray lines of O\,{\sc viii} and C\,{\sc vi}, which are important for estimation of the $\sim$500 MeV photon propagation. 
In the low-ionization zone at larger distances, hydrogen Ly lines and recombination continuum dominate the photon flux with the additional 
strong lines from C\,{\sc iv}, C\,{\sc v}, He\,{\sc i}, He\,{\sc ii}, as well as helium recombination continua. 

\begin{deluxetable}{lccc}
\tabletypesize{\scriptsize}
\tablecaption{BLR Strongest Lines and Recombination Continua Causing 
Jumps in the $\gamma$-Ray Opacity \label{tbl:blr}}
\tablewidth{0pt}
\tablehead{
\colhead{Feature} & \colhead{$\lambda_{\rm BLR}$\tablenotemark{a}} & \colhead {$E_{\rm BLR}$\tablenotemark{b} }  & \colhead{ $E_{\gamma}$\tablenotemark{c} }  \\
            &   ({\AA})   &    (eV)                                     &   (GeV)                                         } 
\startdata
\multicolumn{4}{c}{Low-ionization lines}   \\
O\,{\sc vii}  (blend) & 22 & 560 &  0.47 \\
C\,{\sc v} (blend) & 40.5 & 305 &  0.86 \\
\multicolumn{4}{c}{Low-ionization H\,{\sc i} 10 eV complex}   \\
Ly continuum & 911 & 13.6 &  19.2 \\
Ly$\alpha$  & 1215 & 10.2 &  25.6 \\
C\,{\sc iv} 	              & 1549 & 8.0 & 32.6 \\
\multicolumn{4}{c}{Low-ionization He\,{\sc i} 20 eV complex}   \\
He\,{\sc i} rec. continuum & 504.2 & 24.6 & 10.6 \\
He\,{\sc i} & 584.3 & 21.2 & 12.3 \\
\multicolumn{4}{c}{High ionization lines}   \\
O\,{\sc viii} & 16.01 & 774 &  0.34 \\
O\,{\sc viii} & 18.97 & 653 & 0.40  \\
C\,{\sc vi}  & 33.74 &  367 &  0.71 \\
\multicolumn{4}{c}{He\,{\sc ii} 50 eV complex}   \\
He\,{\sc ii} Ly continuum & 227.8 & 54.4 & 4.8 \\
Fe\,{\sc xv} & 284.2 &  43.6 &   \\
Si\,{\sc xi}  & 303.3 & 40.9 &   \\
He\,{\sc ii} Ly$\alpha$  & 303.8 & 40.8 & 6.4 
\enddata
\tablenotetext{a}{Wavelength of the spectral feature. }
\tablenotetext{b}{Energy of the spectral feature $E_{\rm BLR}$(eV)$\approx$ 12393$/\lambda_{\rm BLR}$(\AA). }
\tablenotetext{c}{Energy of the jump in the $\gamma$-ray opacity $E_{\gamma} (\mbox{GeV})$ $\approx$ $261/E_{\rm BLR}$ (eV) $\approx$ $\lambda_{\rm BLR}$(\AA)/47.5. }
\end{deluxetable}

\subsection{Absorption of GeV Photons} 
\label{sec:abs}

Photons produced in the BLR  constitute the target for the GeV photons propagating through them. 
If the background isotropic photons have strong line-like features peaking at energy $E_0$, 
the photon--photon pair production cross section $\sigma_{\gamma\gamma}(s)$ for a photon of energy $E$ is the function 
of the product of two energies $s=E E_0/(m_{\rm e} c^2)^2$ \citep[see, e.g.][]{GS67,ZDZ88}. 
It has a threshold $s=1$, rapidly grows to the maximum of about 20\% of the Thomson cross section $\sigma_{\rm T}$ at $s\approx 3$, and then slowly decreases. 
The optical depth  for a photon of energy $E$ through the region of size $R$ filled with photons of column density $N_{\rm ph}$ is 
$\tau_{\gamma\gamma}(E,E_0) =  \tau_{\rm T} {\sigma_{\gamma\gamma} (s)}/{\sigma_{\rm T}} $, where 
\begin{equation} \label{eq:taut}
\tau_{\rm T} =  N_{\rm ph}  \sigma_{\rm T} =  \frac{L  \sigma_{\rm T}}{4\pi R c E_0} = 110 \frac{L_{45}}{R_{18}} \frac{10\ {\rm eV}}{E_0} ,
 \end{equation} 
and $L$ is the line luminosity (we defined $Q=10^xQ_x$ in cgs units). 
Equation (\ref{eq:taut}) implies that it is not the luminosity, but its ratio to the typical size (i.e. compactness) 
that determines a role of a specific line in absorption of the $\gamma$-rays. 
Because of the luminosity dependence of the BLR size, the opacity also depends on it as 
\begin{equation} 
\tau_{\rm T} \propto L^{1/2}.
\end{equation}
This provides a  simple 
explanation of why the GeV breaks are observed only in the brightest blazars. 
 
The spectrum transmitted through the region is attenuated as $\propto \exp(-\tau_{\gamma\gamma}(E,E_0))$.
If the incident spectrum is a power law,  a break appears above the threshold energy $(m_{\rm e} c^2)^2/E_0$, where 
the spectral index changes by
\begin{equation} \label{eq:deltagamma}
\Delta\Gamma=- \frac{{\rm d} \ln  \exp(-\tau_{\gamma\gamma}(E,E_0))}{{\rm d} \ln E} \approx  
\max \frac{\sigma_{\gamma\gamma}(s)}{\ln s } \approx \frac{\tau_{\rm T} } {4}.
\end{equation}

 \begin{deluxetable*}{llccccccccc}
\tabletypesize{\scriptsize}
\tablecaption{Spectral Properties of Blazars \label{tbl:data}}
\tablewidth{0pt}
\tablehead{                                    
\colhead{Object} & \colhead{$z$}  &  Power Law & \multicolumn{4}{c}{Broken Power Law} &  \multicolumn{4}{c}{Power Law + Double Absorber} \\
 & & \colhead {$\chi^2$} & 
 \colhead {$\Gamma_1$ }  & \colhead {$\Gamma_2$ } & \colhead {$E_{\rm break}(1+z)$(GeV) } & \colhead {$\chi^2$}  &  
\colhead {$\Gamma$} & \colhead {$\tau_{\rm He}$ }  & \colhead {$\tau_{\rm H}$  }  & \colhead {$\chi^2$}  }           
\startdata
3C 454.3           & 0.859 & 117 &  2.36$\pm$0.02  & 3.60$\pm$0.22  & 4.5$\pm$0.5   & 6.5   &  2.37$\pm$0.02 &  6.1$\pm$0.9    & 18.5$_{-7}^{+19}$   &  4.1  \\
PKS 1502+106  & 1.839 &  55 & 2.15$\pm$0.03  & 2.87$\pm$0.16  & 7.8$\pm$1.5  & 7.8    &  2.13$\pm$0.03 &  1.6$\pm$0.6    & 8.4$\pm$1.6              & 6.3 \\      
3C 279              & 0.536 &  18 &  2.17$\pm$0.07  & 2.56$\pm$0.09  & 1.8$\pm$0.6  & 4.6  &  2.28$\pm$0.04 &  2.0$\pm$1.1    & 4.5$\pm$3.1              &  10.1  \\   
PKS 1510--08   & 0.36   &  13 &  2.43$\pm$0.05  & 2.84$\pm$0.27  &  3.1$\pm$1.8 & 6.6   & 2.45$\pm$0.04 &  2.7$\pm$1.5    & 2.7$_{-2.7}^{+8}$       & 8.1 \\
3C 273              & 0.158 & 10 &   2.82$\pm$0.06  & 3.40$\pm$0.42  &  1.9$_{-1.9}^{+1.0}$ & 6.1 & 2.87$\pm$0.05 &  3.6$_{-3.6}^{+6}$ & 0$_{-0}^{+\infty}$  & 7.8  \\
PKS 0454--234 & 1.003 &  50 &  2.04$\pm$0.05  & 2.81$\pm$0.17  & 5.3$\pm$1.0  & 12.3 & 2.04$\pm$0.04 & 3.0$\pm$0.8      & 9.5$\pm$2.7              &  13.7 \\
PKS 2022--07   & 1.388 &  15 &  2.45$\pm$0.05  & 3.02$\pm$0.17  & 9.6$\pm$4.3  & 11.6 & 2.48$\pm$0.06 & 0.8$_{-0.8}^{+0.9}$   & 2.9$_{-1.8}^{+4.3}$  & 12.9 \\
TXS 1520+319  & 1.487 &  11 &  2.49$\pm$0.07  & 2.89$\pm$0.24  & 4.7$\pm$0.5  & 7.9    & 2.48$\pm$0.74 &  1.7$\pm$1.6     & 6.5$_{-5}^{+9}$      &   7.2 \\
RGB J0920+446 & 2.19  &  21 &  1.99$\pm$0.08  & 3.47$\pm$0.4     & 19$\pm$5    & 7.8     & 2.01$\pm$0.07 &  0$_{-0}^{+0.5}$   & 7.6$\pm$2.9               &  11.9
\enddata
\tablecomments{The number of degrees of freedom is 12 for the power law model and 10 for other models. }
\end{deluxetable*}

The efficiency of $\gamma$-ray absorption within a BLR depends on the photon distribution, with  
the averaged photon--photon pair production cross section  being 
\begin{equation}
\overline{\sigma}_{\gamma\gamma} (E)  = \frac{1}{N_{\rm ph}} \int \sigma_{\gamma\gamma}  (s) N_{\rm ph}(E_0) {\rm d} E_0.  
\end{equation}
Figure~\ref{fig:blr}(b) presents the photon--photon pair production cross section  averaged over the BLR spectra shown in Figure~\ref{fig:blr}(a). 
If the ionization is high, the dominant photon source is the He\,{\sc ii} complex at 40--60 eV (see Table \ref{tbl:blr}), which  produces a break 
in opacity at 4--7 GeV. Above 10 GeV, the opacity is a rather smooth function of energy with additional absorption coming 
from He\,{\sc i}, hydrogen Ly lines and recombination continuum, and C\,{\sc v} $\lambda$2274. In the 0.3--0.7 GeV region the opacity 
has another break due to high-ionization lines of O\,{\sc vii}. Detection of this break requires $\tau_{\rm T}\gtrsim300$, which certainly 
is enough to completely absorb the GeV radiation. 
 
In the low-ionization environment, the strongest absorption is produced by hydrogen Ly lines and recombination continuum, forcing a break in opacity at about 20--30 GeV. At the same time, the break due to helium is also clearly visible at a few GeV. The breaks below 1 GeV would be probably impossible to detect due to low opacity at these energies. The opacity is nearly flat up to 1 TeV \citep{TM09} because of the contribution from additional lines: Mg\,{\sc ii}  $\lambda$2800, H$\alpha$, and He\,{\sc i} $\lambda$10832.

The breaks at 4--7  and 20--30 GeV are expected for a large range of ionization parameters. 
The total opacity in 5--50 GeV range can therefore be very roughly represented as a sum of opacities resulting from two ``lines''  
at ionization energies of He\,{\sc ii} and H\,{\sc i},  54.4 and 13.6 eV, taken in different proportions.
The opacity in these lines can be characterized by the optical depths  $\tau_{\rm He}$ and $\tau_{\rm H}$, 
which reflect the column density of photons at respective energies via Equation (\ref{eq:taut}).
In this simple approximation, which we will call a double-absorber model,  the breaks appear at 4.8 and 19.2 GeV.

\section{Blazars observed by {\em Fermi}} 
\label{sec:fermi}

We analyze the data for several brightest FSRQs from the sample of 12 objects in Table 1 in \citet{Abdo10_blazars} and  choose the same 180 day interval for easier comparison.
We cannot use the standard software distributed by the LAT team, because it does not contain the required spectral models. Instead, we wrote our own version and check its performance, by comparing its results for the simple broken power law models with those obtained by \citet{Abdo10_blazars} using the standard maximum likelihood analysis tool {\it gtlike}. 
 We excluded three objects for which our  simplified analysis is not adequate: 4C+38.41 and PKS 0528+134, because of severe source confusion, and PKS 1908--201 since it is close to the Galactic bulge and the background is high.  
 

\subsection{Data Analysis} 

We use class 3 (diffuse) photons and impose the cuts at zenith angle $<$105$\degr$ and incident angle to the detector axis $<$60$\degr$. The latter cut excludes counts with very low value of the detector response and restricts variation of the count weight by factor $\sim$2.7. We used \verb|P6_V3_DIFFUSE|  version of the response function. The background was measured in a circle with the radius 6$\degr$, separated from the object by $\sim$15$\degr$ and avoiding local sources. We accumulate counts in the circle centered at the source location with the energy-dependent radius $r=\max\{10\degr (E/100\mbox{MeV})^{-0.7}, 0.6\degr\}$ (which corresponds to 95\% containment).

At this stage, we are not interested in the absolute normalization of the spectrum, but only in its shape, therefore our strategy is to sample the response function at photon arrival times (rather than integrating it over time). As the energy resolution in the range of interest (1--30 GeV) is good enough, we can treat the response matrix as diagonal and just prescribe to each photon the weight inversely proportional to the effective area for the given energy and incident angle. We restrict the spectral fit to energies above 200 MeV.
We use binned fitting as it provides us with a very simple and clear indicator of the fit quality, $\chi^2$. 
For a few bins at high energies, where the number of photons is low, we use Poisson likelihood adding $-2\log P(n,\mu)$ to $\chi^2$ (here $n$ is the number of counts in the bin and $\mu$ is the prediction of the model). The number of such bins is small and the  meaning of $\chi^2$ is not significantly affected. For the minimization we use the standard code {\sc minuit} from the CERN library.

 \begin{figure}
\centerline{\epsfig{file= 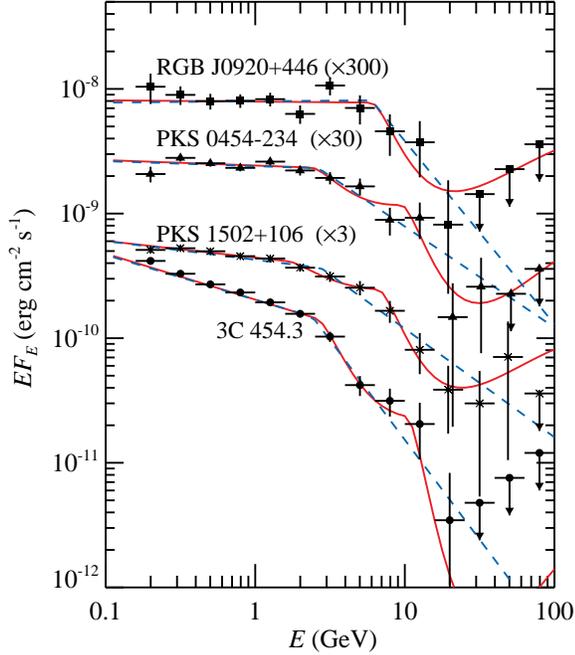,width=7.5cm}} 
\caption{Spectral energy distribution of a few blazars as observed with {\em Fermi}/LAT. 
The  best-fit broken power law and  a power law with the double-absorber models are shown by  the
dashed and solid lines, respectively. 
}	
\label{fig:agn}
\end{figure}

\subsection{Spectral Fits} 

First, we fit the data with a simple power law model. For five objects out of nine from our sample, the fits are acceptable ($\chi^2<18$ for 12 degrees of freedom). 
Next, we apply the broken power law model with parameters $\Gamma_1, \Gamma_2$, $E_{\rm break}$  and free normalization
(see Table \ref{tbl:data}). There is a good agreement with the results by \citet{Abdo10_blazars}, except for RGB J0920+446 (where the best-fit parameters given by \citet{Abdo10_blazars} evidently do not match the spectrum of the object presented in the same paper) and for PKS 0454--234 (where our minimization routine probably found a different local minimum). The breaks  are not statistically significant in PKS 1510--08, 3C 273, PKS 2022--07, and TXS 1520+319. 

 Finally, we have fitted the data with a power law and the double-absorber model  taking 
optical depths $\tau_{\rm He}$ and $\tau_{\rm H}$,  photon index  $\Gamma$  and normalization as free parameters. 
The ``lines'' were redshifted by the appropriate $1+z$ factor. 
The fits  give statistically significant detections of absorption with a good $\chi^2$ for  3C 454.3, PKS 1502+106, PKS 0454--234, and RGB J0920+446 (see Table \ref{tbl:data} and Figure \ref{fig:agn}). 
Upper limits on absorption for PKS 1510--08,  PKS 2022--07, and TXS 1520+319 are significantly smaller than the absorption optical depth for 3C 454.3. There are no significant constraints for 3C 273.  The quality of the fits with the absorption model is about the same as with the broken power law model with the same number of parameters.  In fact, the absorption model is less flexible as the energy, where the power law spectrum breaks, is fixed.

\section{Discussion and summary}

The GeV breaks observed in blazars are well described by $\gamma$-ray absorption via photon--photon pair production on 
He\,{\sc ii} and H\,{\sc i} recombination continuum photons. 
In RGB J0920+446, the absorption is seen only at high energies with the break energy $E_{\rm break}$ corresponding 
to the pair-production threshold on hydrogen recombination photons, 
while in other cases the break is close to the threshold for the absorption on He\,{\sc ii}  recombination continuum. 
The exact position of the break depends on the ionization parameter 
that determines the contribution of metals and affects the position of the centroid of the 50 eV complex. 
If $\tau_{\rm He}$ is small, then the break shifts to 19 GeV as observed in RGB J0920+446.

A rather large ratio of the fitted optical depths  $\tau_{\rm He}/\tau_{\rm H}\sim 1/4$, implies that  
the $\gamma$-ray emitting region has to lie within the high-ionization zone of the BLR with $\log \xi>2$. 
For the brightest $\gamma$-ray object in our sample, 3C 454.3, with its accretion luminosity of about $10^{47}$ erg s$^{-1}$ \citep{SEB88}, 
the high-ionization zone should be within about 0.1 pc.   
This corresponds to about $10^3$ Schwarzschild radii for an $\sim10^9 M_{\odot}$ central black hole \citep{Bonnoli10}.
At such a distance, the  luminosity in the 50 eV complex can be as small as  $\sim$10$^{44}$ erg s$^{-1}$, a per mille of the accretion luminosity, to provide the necessary opacity with $\tau_{\rm He}\sim 6$ (see Equation \ref{eq:taut}).

The opacity measured above 20 GeV and the  Ly$\alpha$ luminosity of 10$^{45}$ erg s$^{-1}$ observed in 3C 454.3 \citep{Wills95}
allow us to estimate the Ly$\alpha$ emission zone size. Taking the recombination continuum luminosity equal to 
that of Ly$\alpha$ and using Equation (\ref{eq:taut}) we get $R_{\rm Ly\alpha}=2\pm1$ pc, which indicates  that 
absorption at these energies happens at a larger distance than absorption by   He\,{\sc ii} photons.

The constraints obtained on the $\gamma$-ray emission site imply that the jet is already accelerated to a relativistic velocity within a thousand 
gravitational radii. It also strongly  constrains the  mechanisms for $\gamma$-ray production. 
The possible sources of soft photons for Comptonization in the jet are the accretion disk \citep{DS93} and the BLR \citep{SBR94}.
As a source of photons, the dust emission at 10 pc scale \citep{BSM00,SMM08} cannot be important. 

Let us also remark that the GeV photons absorbed in the BLR produce electron--positron pairs which,  
spiraling in the magnetic field, radiate away their energy isotropically. 
This emission cannot compete with the beamed emission from the jet, but can contribute to the high-energy emission of radio galaxies 
observed at large angles to the jet axis \citep[see, e.g.][]{RB10}.

Our interpretation of the GeV breaks implies that additional breaks (depressions) at 0.3--0.7 GeV 
produced by the soft X-ray lines within the high-ionization zone should be seen, once the  photon statistics is high  enough. 
The $\gamma$-ray spectroscopy can be used as a powerful tool for studying the extreme-UV and soft X-ray emission in the 
quasars' vicinity, which is otherwise hidden from us by interstellar absorption.

\acknowledgments
This research was supported  by the Academy of Finland grants 127512 and 133179 and the V\"ais\"al\"a foundation.
The research made use of public data obtained from the {\it Fermi} Science Support Center.
We thank Markus Boettcher, Evgeny Derishev,  Julian Krolik, Amir Levinson, Fabricio Tavecchio, and Dmitry Yakovlev for useful comments and 
Tim Kallman for his help with {\sc xstar}. 




\end{document}